\documentclass[journal]{IEEEtran}
\usepackage{graphicx}
\graphicspath{ {photo/} }
\usepackage{xcolor,soul,framed} %,caption

\colorlet{shadecolor}{yellow}
\usepackage[cmex10]{amsmath}
\usepackage{array}
\usepackage{mdwmath}
\usepackage{mdwtab}
\usepackage{eqparbox}
\usepackage{url}
\usepackage{float}
\usepackage{caption}
\usepackage{wrapfig}
\usepackage{graphicx}
\usepackage{subfig}
\begin{document}
% \bstctlcite{IEEEexample:BSTcontrol}
    \title{Comprehensive Analysis and Improvements in Pansharpening Using Deep Learning}
  \author{Mahek Kantharia,
      Neeraj Badal,
      Zankhana Shah
     }

\maketitle
% === ABSTRACT ====================================================================
% =================================================================================
\begin{abstract}
%\boldmath
Pansharpening is a crucial task in remote sensing, enabling the generation of high-resolution multispectral images by fusing low-resolution multispectral data with high-resolution panchromatic images. This paper provides a comprehensive analysis of traditional and deep learning-based pansharpening methods. While state-of-the-art deep learning methods have significantly improved image quality, issues like spectral distortions persist. To address this, we propose enhancements to the PSGAN framework by introducing novel regularization techniques for the generator loss function. Experimental results on images from the Worldview-3 dataset demonstrate that the proposed modifications improve spectral fidelity and achieve superior performance across multiple quantitative metrics while delivering visually superior results. 
\end{abstract}

% === KEYWORDS ====================================================================
% =================================================================================

% === I. INTRODUCTION =============================================================
% =================================================================================
\section{Introduction}

\IEEEPARstart
Owing to payload, cost and data constraints, remote sensing satellites are unable to capture images having both high spatial and spectral resolution.Therefore satellites provide two types of images  a high spatial resolution panchromatic image and a high spectral-resolution multispectral image. Both these images are taken by sensors looking over the same area at the same angle and hence requires no registration task. Pansharpening or Panchromatic sharpening aims to combine these two images to achieve higher spatial resolution while preserving specific spectral attributes.

Over the years, many algorithms have addressed this task. These techniques are mailnly divided into component substitution, multiresolution analysis, sparse representation and variational approaches. While these traditional techniques are relatively easy, they face several difficulties such as spectral distortions, spatial detail injection limitations, and limitations based on theoretical assumptions. Performance of these methods is also variable across different sensor data, and land cover characteristics. 

With recent advances made by deep neural networks for
image processing applications, researchers have also explored the avenue for pan-sharpening. Deep learning has become a popular technique for image processing tasks, including pansharpening, due to its ability to learn complex nonlinear mappings between input and output images. In the context of pansharpening, deep learning can be used to learn the complex relationships between low-resolution multispectral images and high-resolution panchromatic images. Unlike traditional methods that rely on handcrafted features and linear models, deep learning can automatically learn the most discriminative features from the input data, and use them to generate high-quality output images. 

Many researchers have used deep-learning for pansharpening. Inspired by image super-resolution \cite{srcnn}, Masi et al. \cite{pnn} constructed a three-layer convolutional neural network for pan-sharpening. \cite{pannet} trains ResNet architecture in high-pass domain and adds MS input to it's output.\cite{PSGAN} uses a Conditional GAN to generate pansharpened image. Even though these methods are a great improvement over traditional methods they still show spectral distortions. In this paper, we propose a perceptual loss function that reduces this spectral distortion. We hypothesize that by minimizing the distance between the pansharpened and original MS in the high-level feature subspace, a more spectrally accurate image can be generated. To do this we calculate the gram matrix of the output of the bottom-most layer of the generator from PSGAN and use L1 loss between the two as an enriching term in the generator loss function.

This paper is organized as follows. In section II various traditional as well as deep-learning based pansharpening methods are briefly reviewed. Section III presents the methodology used to improve pan-sharpening. Experimental results and comparisons are provided in Section IV. Conclusions are drawn in Section V.

\section{Related Work}
This section briefly reviews Component substitution, Multiresolution analysis and Deeplearning based pansharpening methods

\subsection{Component Substitution}

Component Substitution methods are based on projection of the multispectral images into a space, where the spatial and spectral components can be separated. The spatial component is then replaced by the panchromatic image. Since greater correlation between the replaced component and the panchromatic image leads to lesser distortion, the panchromatic image is histogram matched with the replaced component. The process is completed by doing inverse transformation to bring back the multispectral image to the original space.

% \begin{figure}
%     \centering
%     \includegraphics[width=0.6\linewidth]{photo/CS methods.png}
%     \caption{Component Substitution methods}
%     \label{fig:enter-label}
% \end{figure}

Some of the techniques belonging to this class are IHS, GIHS, Brovey transform, PCA, Gram-Schmidt analysis. 

\subsubsection {IHS \cite{ihs_cite1,ihs_cite2}, GIHS}
IHS colour model is used as it separates the spatial information (intensity component) from the spectral information (hue and saturation). Hence, it is possible to manipulate the spatial information while keeping the same spectral information.

The hue component describes the color in the form of an angle between 0 to 360 degrees. It is determined by the relative proportions of red, green and blue The saturation describes the purity of the color, how much color is dilated with white light and its value ranges between 0 and 1. The IHS solid adds the dimension of intensity with black at the bottom and white at the top. Shades of gray run along the axis of the solid. 

% \begin{center}
% \includegraphics[width=5cm]{photo/IHS-removebg-preview.png}
% \end{center}
 
IHS yield adequate spatial enhancement but introduces spectral distortion. In \cite{ihs_reason}, it is demonstrated that saturation component changes after changes in the intensity value. The product of intensity and saturation is a constant value. Hence, saturation and intensity are inversely proportional to each other. The new saturation value is expanded if Pan value is less than internsity, \textit I value and it is compressed when Pan value is greater than the \textit I value. Studying the relative spectral response, the authors find that the RGB bands do not fall within the same range of the panchromatic band. 

Furthermore, the response of the Pan is expanded beyond the NIR band. Since the spectral response of Pan and \textit I are bound to change, to reduce the colour distortion, they introduce a generalized IHS method that responds to the NIR band.

\subsubsection{Brovey}
Brovey transform \cite{pansharpening} normalizes the spectral bands before they are multiplied with the panchromatic image. It retains the corresponding spectral features of each pixel and transforms the luminance information into a panchromatic image which then gets replaced by the histogram high resolution panchromatic image.

However, the Brovey transform method assumes that the spectral response of the PAN image represents the overall spectral content of the MS image. However, this assumption may not hold true in all cases, leading to spectral distortion in the sharpened image.
\subsubsection {PCA}
The MS image is taken as an input to principle component analysis procedure. The first principle component represents the maximum variance direction of the data. The Pan data is histogram matched with the first principal component. The results of which are used to replace the first principle component and the data is retransformed back to its original space. 
The justification used to replace the first component is that the first principle component will have information which is common to all bands which is the spatial information while spectral information unique to any of the bands is mapped to the other components. However some of the spatial information may not be mapped to the first component, depending on the degree of correlation and spectral contrast existing among the MS bands \cite{pansharpening}.

\subsubsection{Gram-Schmidt analysis}
The Gram-Schmidt (GS) algorithm is commonly used in remote sensing to orthogonalize matrix data or bands of a digital image. This process removes redundant or correlated information that is contained in multiple bands, resulting in a more accurate outcome.

The multispectral (MS) bands are resampled or interpolated to the same scale as the panchromatic (PAN) band.A lower-resolution panchromatic (PAN) band is simulated and used as the first band of the input to the GS transformation. Then each MS band vector is projected onto the (hyper)plane established by the previously determined orthogonal vectors. Pansharpening is accomplished by replacing the first vector with the histogram-matched PAN before the inverse transformation is performed.
Two methods are used for creating a low resolution PAN \cite{pansharpening}. In the first method the LRMS bands are combined into a single lower-resolution PAN (LR PAN) as the weighted mean of MS image. The second method simulates the LR PAN image by blurring and subsampling the observed PAN image. However, the first method suffers from spectral distortion and the second method sufferes from low sharpeness. In order to avoid this drawback an enhanced GS method is used, where the LR PAN is generated by a weighted average of the MS bands and the weights are estimated to minimize the MMSE with the downsampled PAN. 

GS is a generalization of PCA in which the first principle component can be choosen and the other components are made to be orthagonal to one another and the first component.

GSA assumes that the orthogonalized multispectral bands preserve the original spectral information. However, the orthogonalization process can lead to spectral distortion in the sharpened image. This distortion may result in colour shifts, an inaccurate representation of the original spectral content, or the introduction of artificial spectral artefacts. Also, GSA prioritises the enhancement of spatial resolution by utilising the panchromatic image. However, this can come at the expense of spectral resolution, potentially leading to a loss of fine spectral details in the sharpened image.

\subsection{MRA}
Another class of methods are Multiresolution analysis based methods that aim to extract the spatial information (high-frequency detail) from the PAN image by wavelet transform, Laplacian pyramid, etc., in the first step and then inject it to the up-sampled MS images to generate the fused image

\subsubsection{HPF}
A high pass filter is used to obtain high-pass information from the Pan image. The HPF results are added pixel by pixel to the lower spatial resolution MS image

\subsection{Deeplearning based methods}
While these traditional techniques are relatively easy, they face several difficulties such as spectral distortions, spatial detail injection limitations, and limitations based on theoretical assumptions. Performance of these methods is also variable across different sensor data, and land cover characteristics. 

With developments in machine learning (ML) and deep learning (DL) in the last decades, these technologies started to be widely used in image processing, such as image classification, image segmentation, object
detection super-resolution, pan-sharpening and reconstruction.
\subsubsection{PNN}
Researchers in \cite{pnn} have build upon architecture proposed in \cite{srcnn} for super-resolution problem and converted it to solve the pansharpening problem by leveraging to it the huge domain-specific knowledge available. 
The 4-band Multispectral components are upsampled and interpolated and are stacked with the panchromatic band to form the 5-component input. 
In addition to this, the authors add more planes corresponding to some well-known radiometric indices.

Mean Square error between the pansharpened image and its reference is used as the loss function

\begin{equation}\label{loss function}
    L(W) = \frac{1}{n} \sum_{n=1}^{n}||(M_n-\hat{M_n}(W))||_2
\end{equation}

where $M_n$ is the reference image, $\hat{M_n}$ is the pansharpened image and $n$ is the number of batch size

\subsubsection{PanNet}
The design of PNN is relatively simple and needed to be improved. But deep neural networks are difficult to optimize. \cite{resnet} demonstrated the same problem and devised a clever solution that allowed the layers to learn residual functions with respect to the layer inputs instead of learning the unreferenced functions from scratch. This allowed training over 2000 layers with increasing accuracy. 

\begin{figure}
    \centering
    \includegraphics[width=0.9\linewidth]{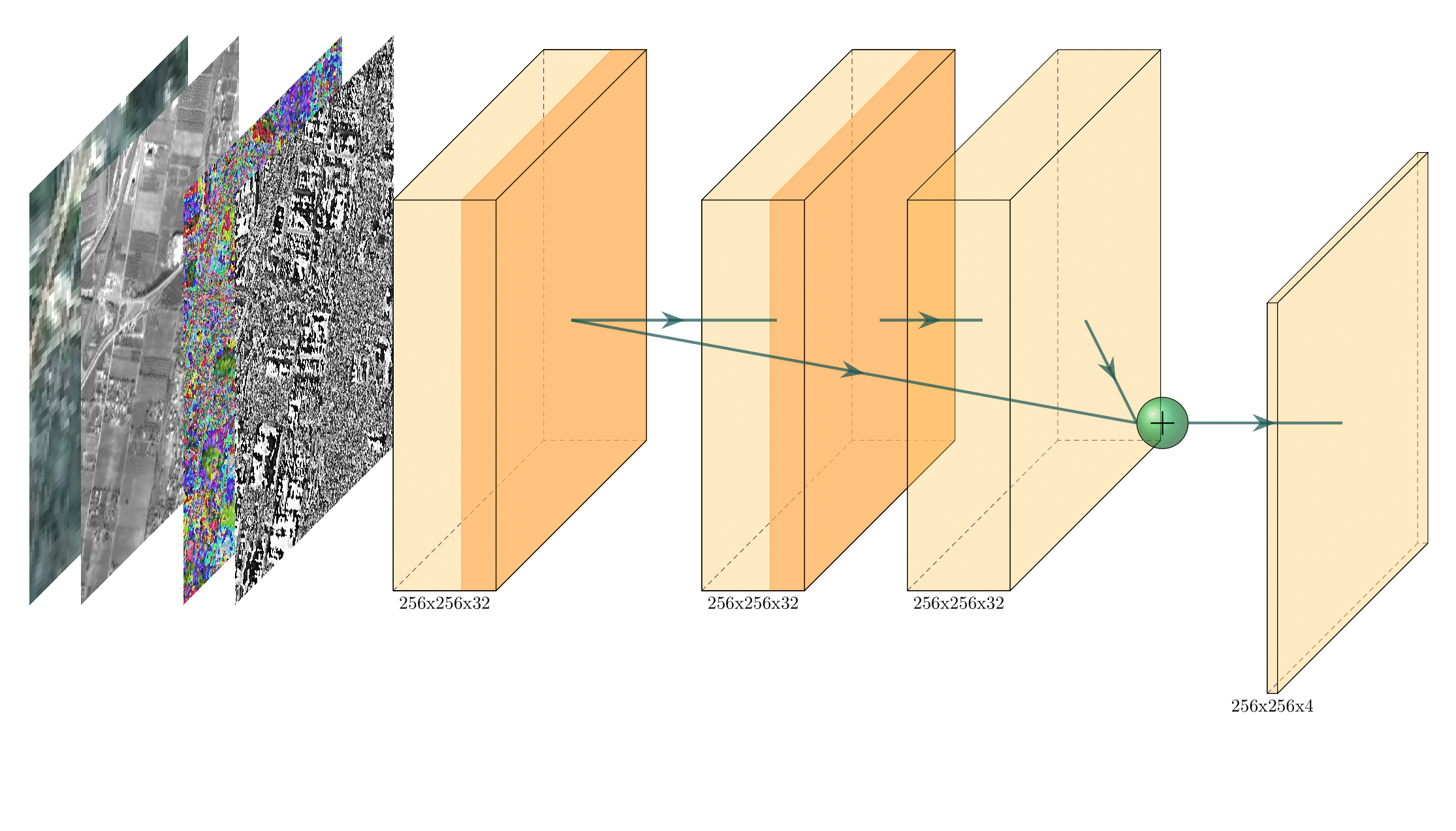}
    \caption{PanNet model}
    \label{fig:enter-label}
\end{figure}

% \includegraphics[width=0.9\linewidth]{photo/pannet.png}
% \caption{PanNet}
% \end{center}

This approch was implemented in the task of Pansharpening by the authors of \cite{pannet} and is called PanNet. The researchers trained the ResNet in the high-pass domain to preserve spatial features and simply added the upsampled MS input to the model output to preserve spectral features.

PanNet uses the same loss function as PNN.

\subsubsection{PSGAN}
Both the approaches above used Euclidean distance between the predicted and reference image as a loss function which would cause blurring effects. 

Another important breakthrough in the DL field is Generative Adversarial Networks \cite{gans} where a generative model tries to generate an image like the real image and is pitted against an adversary: a discriminative model that learns to determine whether an image is generated or real. This later lead to \cite{condgans} where Conditional GANs have been used to for image to image translation.  

In \cite{PSGAN} a Generative Adversarial Network(PSGAN) was first applied for pansharpening. This network consisted of a two-stream input generator inspired by the U-NET \cite{unet}architecture and a fully convolutional discriminator similar to \cite{condgans}. The work also demonstrated that l1 loss produced better results than l2 loss.

The loss function of the generator and the discriminator are:
\begin{figure*}[h]
\centering

\includegraphics[width=0.9\linewidth]{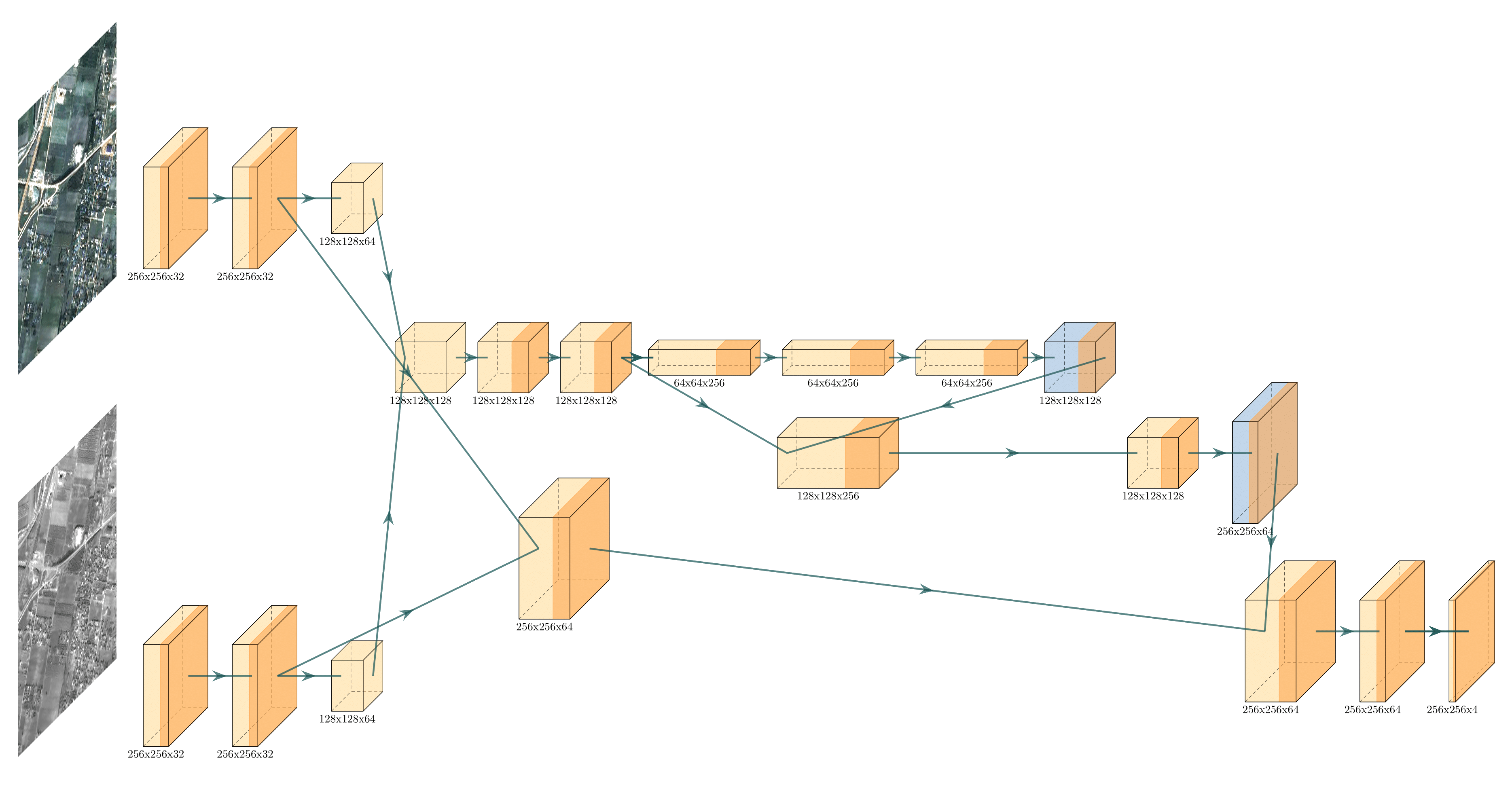}
\caption{PSGAN Generator network} 

\end{figure*}

% \begin{figure*}[h]
% \centering\includegraphics[width=0.9\linewidth
% ]{photo/FCN8__Copy___6_-1-removebg-preview.png}
% \caption{PSGAN generator} 
% \end{figure*}
\begin{center}
\begin{equation}\label{loss_func_gen}
\begin{aligned}
    L(G) = \frac{1}{n} \sum_{n=1}^{n}[-\alpha \log D_{\theta_D}(M_n,G_{\theta_D}(M_n,P_n)) \\ + \beta ||\hat{M_n}-G_{\theta_D}(M_n,P_n)||_1]
\end{aligned}
\end{equation}
\end{center}

\begin{figure*}[h!]

 \centering\includegraphics[trim=0cm 0cm 2cm 0cm,clip=true,width=0.9\linewidth]{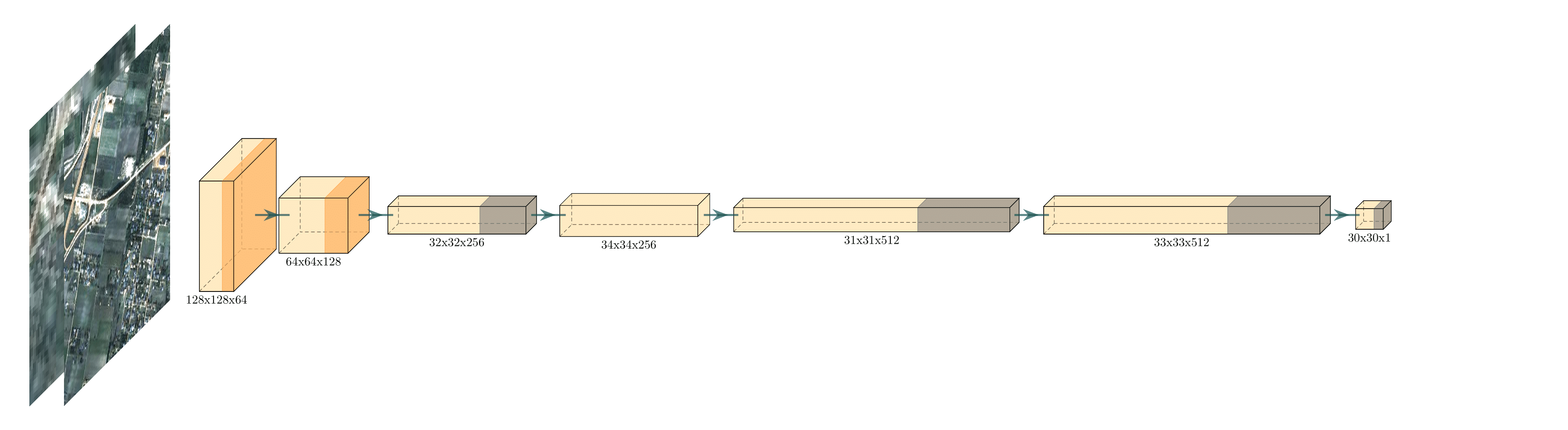}
 \caption{PSGAN discriminator network} 
\end{figure*}

\begin{center}
\begin{equation}\label{loss_func_disc}
\begin{aligned}
    L(D) = \frac{1}{n} \sum_{n=1}^{n}[1-\log D_{\theta_D}(M_n,G_{\theta_D}(M_n,P_n)) \\ + \log D_{\theta_D}(M_n,\hat{M_n})]
\end{aligned}
\end{equation}
\end{center}

where $M_n$ is the LRMS image, $P_n$ is the HRPAN image, $\hat{M_n}$ is the HRMS image and G and D are the Generator and Discriminator respectively.

\section{Proposed methodology}
Our method builds upon PSGAN. Apart from the L1 loss between $\hat{M_n}$ and $G_{\theta_D}(M_n,P_n)$ we propose three new loss functions:
\subsection{Loss function based on SAM}
In order to reduce the spectral distortions seen in PSGAN method, we devised a loss function like SAM (Spectral Angular Mapper) \cite{sam} and used it as a regularizing term is the generator loss function.

\begin{center}
\begin{equation}\label{sam_loss}
\begin{aligned}
    Sam\ loss(\widetilde M_n,\hat{M_n}) = 1-\frac{\sum_{n=1}^{n}\widetilde M_n \cdot \hat{M_n}}{\sum_{n=1}^{n}{\widetilde M_n}^2\sum_{n=1}^{n}\hat{M_n}^2}
\end{aligned}
\end{equation}
\end{center}
where $\widetilde M_n$ is output of $ G_{\theta_D}(M_n,P_n)$.
\subsection{Sam loss on both resolutions}
We applied the above loss function on both reduced and original resolutions and used it as a regularizer in the generator loss function.

\begin{center}
\begin{equation}
\label{sam_loss_multires}
\begin{aligned}
    Total\ Sam\ loss= 0.5 \times Samloss(\widetilde 
    M_n,\hat{M_n})+ \\
    0.5 \times Samloss(\widetilde M_{down_n}, M_n)
\end{aligned}
\end{equation}
\end{center}

where $\widetilde M_{down_n}$ is obtained by downsampling $\widetilde M_n$ by r.
\subsection{Perceptual loss}

We created a perceptual loss function that reduces this spectral distortion. We hypothesize that by minimizing the distance between the pansharpened and original MS in the high-level feature subspace, a more spectrally accurate image can be generated.

In order to generate the high-level feature subspace, we take a network with same architecture as the generator and add dropout layers to add noise. We train this network on MS images and take the L2 norm between the generated and input as loss function. 

We take L2 norm between these two high-level feature subspaces taken from the bottleneck layer of the pretrained U-NET as a regularizer to the generator loss function.

\begin{center}
\begin{equation}
\label{perceptual_loss}
\begin{aligned}
Perceptual\ loss= \| U(\widetilde M_n)-U(\hat{M_n})\|_2
\end{aligned}
\end{equation}
\end{center}

\subsection{Gram matrix based perceptual loss}
Instead of directly taking L2 norm like in the previous method, we calculate the gram matrix of the high level features of the generated and the original image and the euclidean distance between them is minimized. 

\begin{equation}\label{gram_matrix_1}
\begin{aligned}
G_{U(\hat{M})}=U(\hat{M})^T \times U(\hat{M})
\end{aligned}
\end{equation}

\begin{equation}\label{gram_matrix_2}
\begin{aligned}
G_{U(\widetilde{M})}=U(\widetilde{M})^T \times U(\widetilde{M})
\end{aligned}
\end{equation}

\begin{equation}\label{gm_based_perceptual_loss}
\begin{aligned}
GM \ based \ Perceptual\ loss = \|G_{U(\hat{M})}-G_{U(\widetilde{M})}\|_2
\end{aligned}
\end{equation}

\subsection{Gram matrix based reconstruction loss}
We create a regularizer which is based on minimizing the distance between gram matrix of the reference and pansharpened patches. We use this loss along with generator loss and give both the same weightage

\begin{equation}\label{reconstruction_loss}
\begin{aligned}
GM \ based \ Reconstruction\ loss = \|G_{\hat{M}}-G_{\widetilde{M}}\|_2
\end{aligned}
\end{equation}

% We choose the activations of the bottleneck layer of the Generator architecture for this purpose.

% The network that generated the the high level features of the original multispectral image (shown below) $\hat{M_n}$ has the loss function:

% \begin{equation}\label{loss_func_disc}
% \begin{aligned}
%     L(P) = GM(\phi(\hat{M_n}) )-GM(\phi(G_\theta_D(M_n,P_n)))    
% \end{aligned}
% \end{equation}
% where GM is the Gram matrix construction function defined by $GM(X)=X^T X$ and $\phi$ represent the bottleneck layer activations

\begin{figure*}[h]
\centering\includegraphics[trim=0cm 0cm 0.5cm 0cm,clip=true,width=0.9\linewidth
]{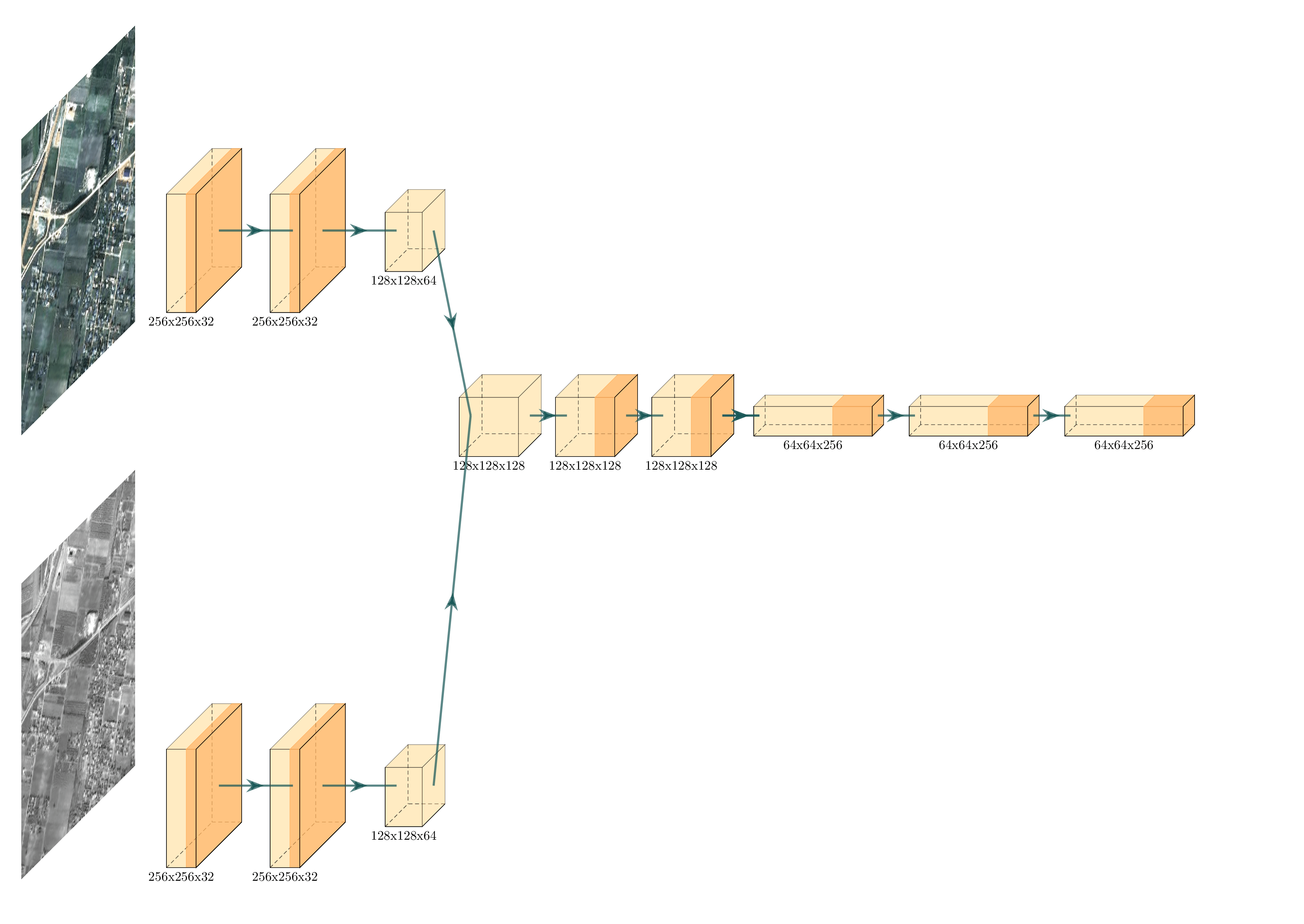}
\caption{Network to generate high level features of multispectral images} 
\end{figure*}

In all of the above cases we use the loss functions as regularizers and the final loss function of the generator becomes:
\begin{equation}\label{loss_func_disc}
\begin{aligned}
    L = \eta_1 L(G) + \eta_2 Regularizer()
\end{aligned}
\end{equation}

In order to calculate $\eta_1 \; and \; \eta_2 $ we employ hyperparameter tuning using gradient descent.

\section{Experiments}
\subsection{Dataset and Performance measure}
We train and test our network on dataset acquired from Worldview-3 satellite. The spatial resolution is 0.34m for PAN and 1.38m for MS. The dataset consists of images taken over three cities: Paris, Vegas and Shanghai. Wald's protocol\cite{wald} is followed to downsample both the MS and PAN by a factor of 4 so that the resulting pansharpened image can be compared with the original MS image. Anti-aliasing is used for downsampling as it blurs the patches before the downsampling process. Hence, there is no need to apply a smoothing kernel. The training dataset consists of patches of size 256x256 from the datasets of Paris and Vegas. While the patches from Shanghai are used for testing. 

Popular quantitative Evaluation metrics used are:
SAM: Spectral Angle Mapper \cite{sam}, ERGAS: Global adimens, relative synthasis error \cite{ergas}
Q4: 4 band average universal image quality index[] and \cite{q4}
SSIM: Structural Similarity index measure \cite{ssim}

% \subsection{Comparison with the State of the Art}
% The proposed methods are compared with the state of the art techniques: GIHS, Brovey, PCA, HPF, PSGAN.  

\begin{table*}
\centering    
\begin{tabular}
{||c | c c c c c||} 
 \hline
 Method & SSIM & SAM & ERGAS & Q$2^4$ & QNR \\ [0.5ex] 
 \hline\hline
 GIHS & 0.812915 & 0.098976 & 5.592451 & 0.794797 & 0.938389\\ 
 \hline
 Brovey & 0.808367 & 0.090561 & 5.549419 & 0.785414 & 0.938489\\ 
 \hline
 PCA & 0.638585 & 0.127973 & 8.353011 & 0.554729 & 0.922361\\ 
 \hline
 HPF & 0.636111 & 0.092269 & 7.989891 & 0.659868 & 0.949295\\ 
 \hline
 PSGAN & 0.865000 & 0.105504 & 4.292280 & 0.883587 & 0.954166\\
 \hline
 PSGAN with SAM based loss(B) & 0.871410 & \textbf{0.079872} & 4.189323 & 0.885042 & 0.954286 \\
 \hline
 PSGAN with Gram Matrix based reconstruction loss(E) & \textbf{0.891008} & 0.089297 & \textbf{3.736660} & \textbf{0.904016} & \textbf{0.954592}\\
 \hline
\end{tabular}
\caption{Quantitative comparison for traditional methods and models trained on 13,000 patches}
\end{table*}

\begin{table*}
\centering    
\begin{tabular}
{||c | c c c c c||} 
 \hline
 Method & SSIM & SAM & ERGAS & Q$2^4$ & QNR \\ [0.5ex] 
 \hline\hline
 
 PSGAN & 0.871230 & 0.099752 & 4.069809 & 0.884941 & 0.955353\\
 \hline
 PSGAN with perceptual loss(C) & \textbf{0.876108} & \textbf{0.096551} & \textbf{4.056470} & \textbf{0.887232} & \textbf{0.955276} \\
 \hline
 PSGAN with Gram Matrix based perceptual loss(D) & 0.851393 & 0.110626 & 4.456317 & 0.869226 & 0.954790\\
 \hline
\end{tabular}
\caption{Quantitative comparison for models trained on 500 patches}
\end{table*}

\begin{figure*}
\captionsetup[subfloat]{labelformat=empty}
\setlength{\tabcolsep}{0.5pt}
\renewcommand{\arraystretch}{0}
\begin{tabular}{ccccc}
\subfloat{\includegraphics[width = 0.2\textwidth]{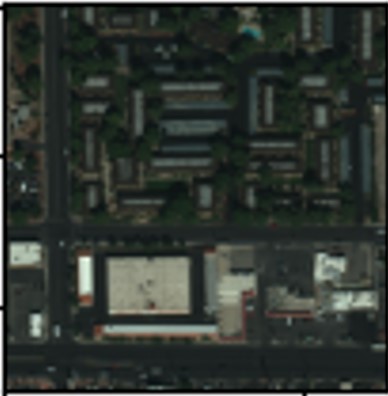}} &
\subfloat{\includegraphics[width = 0.2\textwidth]{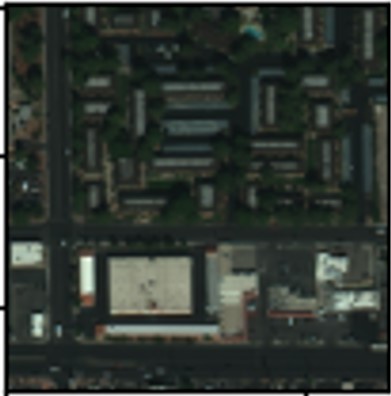}} &
\subfloat{\includegraphics[width = 0.2\textwidth]{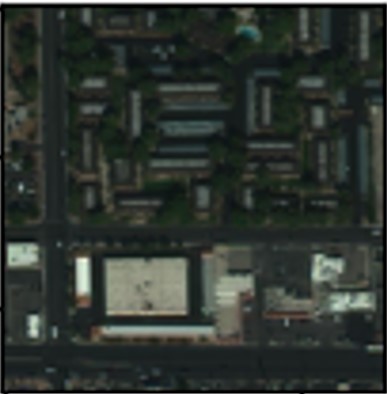}} &
\subfloat{\includegraphics[width = 0.2\textwidth] {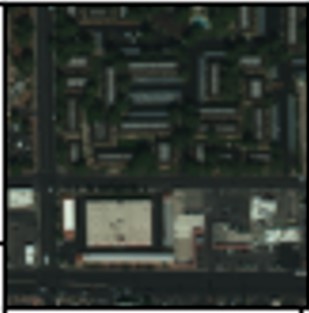}} &
\subfloat{\includegraphics[width = 0.2\textwidth]{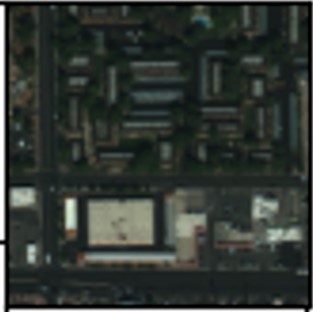}}\\
\subfloat{\includegraphics[width = 0.2\textwidth]{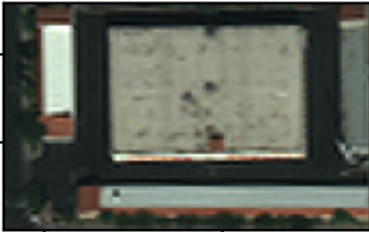}} &
\subfloat{\includegraphics[width = 0.2\textwidth]{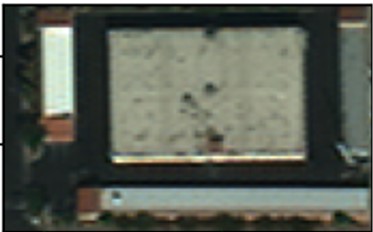}} &
\subfloat{\includegraphics[width = 0.2\textwidth]{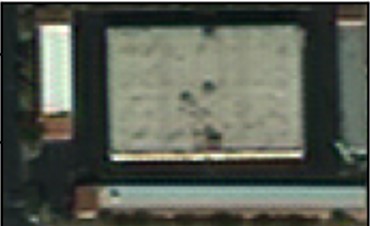}} &
\subfloat{\includegraphics[width = 0.2\textwidth,trim={0 0 0 10pt},clip]{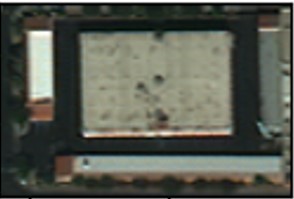}} &
\subfloat{\includegraphics[width = 0.2\textwidth,trim={0 0 0 10pt},clip]{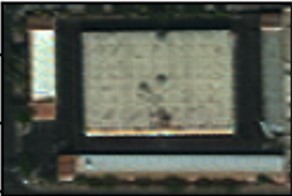}}\\
\subfloat{\includegraphics[width = 0.2\textwidth]{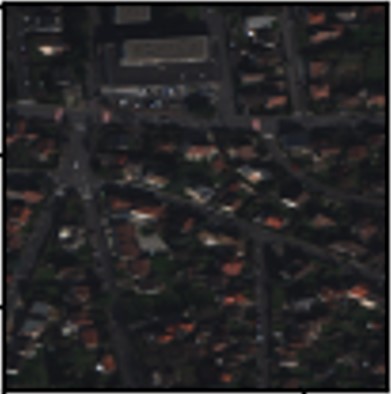}} &
\subfloat{\includegraphics[width = 0.2\textwidth]{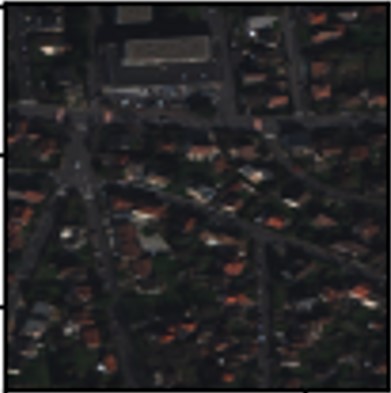}} &
\subfloat{\includegraphics[width = 0.2\textwidth,trim={0 0 0 3pt},clip]{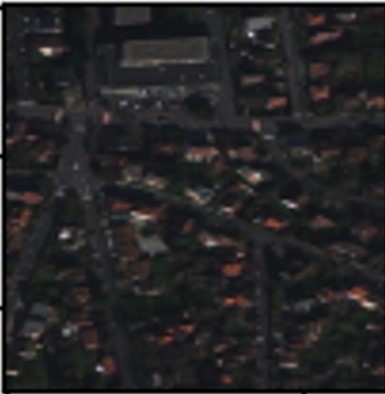}} &
\subfloat{\includegraphics[width = 0.2\textwidth]{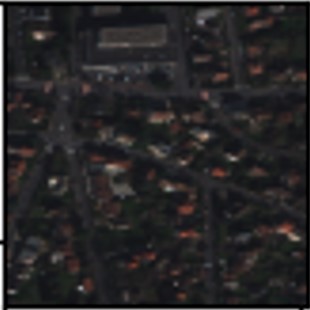}} &
\subfloat{\includegraphics[width = 0.2\textwidth]{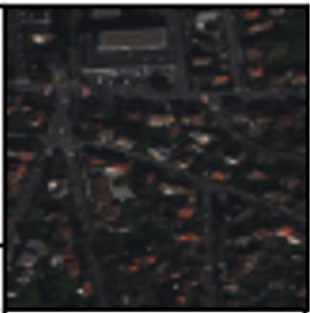}}\\
\subfloat[Ground Truth]{\includegraphics[width = 0.2\textwidth,trim={0 0 0 10pt},clip]{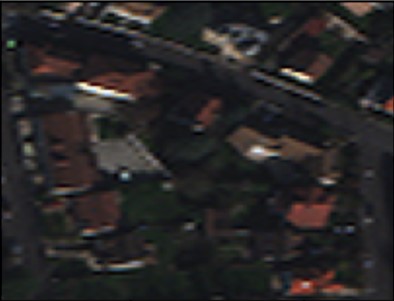}} &
\subfloat[PSGAN]{\includegraphics[width = 0.2\textwidth,trim={0 0 0 10pt},clip]{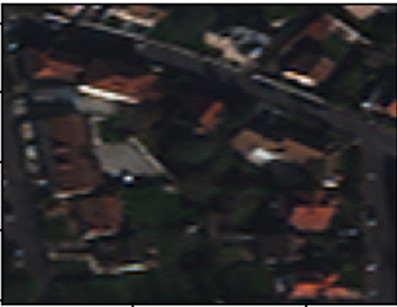}} &
\subfloat[perceptual loss(C)]{\includegraphics[width = 0.2\textwidth,trim={0 0 0 10pt},clip]{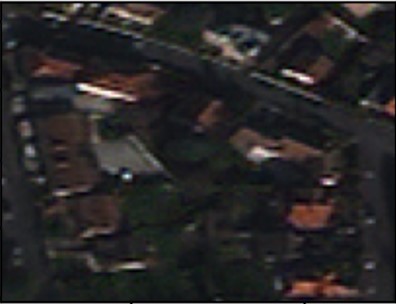}} &
\subfloat[recontruction loss(E)]{\includegraphics[width = 0.2\textwidth]{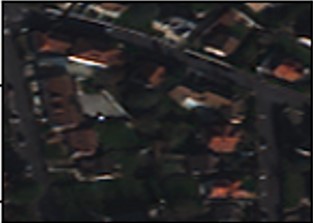}} &
\subfloat[SAM based loss(B)]{\includegraphics[width = 0.2\textwidth]{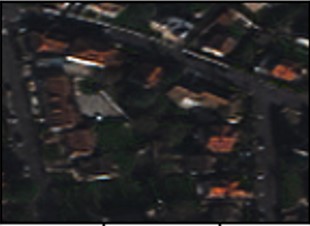}} 

\end{tabular}
\caption{Pansharpening results on Test dataset}
\end{figure*}

\section{Conclusion}
This work introduces several new regularization techniques for the generator loss function in PSGAN. Experimental results demonstrate that the Gram matrix-based reconstruction loss significantly enhances overall performance across most metrics, with the exception of SAM, where the SAM-based loss achieves the best improvement. On smaller datasets, the perceptual loss function shows notable enhancements compared to PSGAN, although the Gram matrix-based perceptual loss leads to a slight performance degradation.

\section*{Acknowledgment}
The authors would like to thank everyone at the IAQD department of SAC, ISRO for their kind cooperation and encouragement. We would also like to thank the department for providing the resources. These resources have played a significant role in enabling us to conduct research, access relevant literature, and acquire necessary datasets for experimentation and analysis.

% or


% ============================================
%\appendices
%Appendix one text goes here %\cite{Roberg2010}.

%\section{}

%\section*{Acknowledgment}

%The authors would like to thank D. Root for the loan of the SWAP. The SWAP that can ONLY be usefull in Boulder...

% % by themselves when using endfloat and the captionsoff option.
% \ifCLASSOPTIONcaptionsoff
%   \newpage
% \fi
\newpage
\bibliographystyle{IEEEtran}
\bibliography{Bibliography}



\end{document}